# Photoluminescence features and nonlinear-optical properties of the Ag$_{0.05}$Ga$_{0.05}$Ge$_{0.95}$S$_2$eEr$_2$S$_3$ glasses

V.V. Halyan[a], V.O. Yukhymchuk[b], Ye.G. Gule[b], K. Ozga[c], K.J. Jedryka[c], I.A. Ivashchenko[d], M.A. Skoryk[e], A.H. Kevshyn[a], I.D. Olekseyuk[d], P.V. Tishchenko[d], M.V. Shevchuk[f], M. Piasecki[g,*] a Department of Experimental Physics and Technologies for Information Measuring, Lesya Ukrainka Eastern European University, 13 Voli Avenue, Lutsk, 43009, Ukraine
b V. Lashkaryov Institute of Semiconductor Physics, NAS of Ukraine, 45 Prospect Nauky, Kyiv, 03028, Ukraine
c Institute of Optoelectronics and Measuring Systems, Faculty of Electrical Engineering, Czestochowa University of Technology, Armii Krajowej 17, Czestochowa, Poland
d Department of Inorganic and Physical Chemistry, Lesya Ukrainka Eastern European University, 13 Voli Avenue, Lutsk, 43009, Ukraine
e Nanomedtech LLC, 68, Antonovycha Str., 03680, Kyiv, Ukraine
f Lutsk National Technical University, 75 Lvivska Street, Lutsk, 43000, Ukraine
g Institute of Physics, J.Dlugosz University in Częstochowa, Armii Krajowej 13/15, Częstochowa, PL-42-217, Poland

ABSTRACT

Preparation technology, the structure determination, multiband luminescence and nonlinear optical properties (NLO) of the chalcogenide glasses are subject of present work. The Ag$_{0.05}$Ga$_{0.05}$Ge$_{0.95}$S$_2$eEr$_2$S$_3$ glass samples with two different 0.12 and 0.27 at.% Er content were prepared by classical two-stage melt-quenching method. Glass state and morphology were confirmed by X-ray and EDS techniques. Influence of the Erbium doping on the luminescence and NLO properties was investigated. Based on the diagram of energy levels for the Er$^{3+}$ ions, we proposed a model that explains photoluminescence (PL) emission mechanism. It should be emphasized that investigated glasses (in comparison with other materials) are exceptional in that all bands of PL are intense due to small energy losses at excitation and emission of ions Er$^{3+}$. The photoinduced second and third harmonic generations have been measured in the reflected geometry and compared with reference – BiB$_3$O$_6$ (BBO): 10%Nd crystal.

## 1. Introduction

Semiconductors doped with rare earth metals (RE) are of considerable interest due to the growing needs of the industry in optoelectronic devices operating in the spectral range compatible with telecommunications windows. The most commonly used RE ion is erbium because possess wide emission band and modest energy losses in fiber optics at 1.5 μm wavelength. Therefore Er-doped materials, both crys-talline and amorphous, are also used as active media in laser technology [1], displays [2], optical amplifiers [3], photonic devices [4], light converters [5], non-contact temperature sensors [6–8] and in devices that operate under γ-radiation [9–11]. In addition, they can be used as multifunctional, nonlinear optical materials in optoelectronics [12–15].

Due to the optical transmission window reaching up to several micrometers, chalcogenide glasses (as opposed to oxide) are particularly predisposed for applications in the infrared range. Phase equilibria and the boundaries of the glass formation region in the reciprocal system AgGaS$_2$ + GeSe$_2$ ⇔ AgGaSe$_2$ + GeS$_2$ were determined and presented in Ref. [16]. Was established that glass matrix with the composition Ag$_{0.05}$Ga$_{0.05}$Ge$_{0.95}$S$_2$ shows the best transparency. Later, were prepared glass samples doped with 0.12 and 0.27 at.% Er and investigated the principal structural units of the glass-forming matrix by Raman spec-troscopy, as well as photoluminescence (PL) spectra under excitation by 532 and 980 nm wavelengths [8,17]. Partial clustering of erbium ions was established from EPR and static magnetization studies, and ad-ditionally the effect of γ-irradiation on PL properties in these glasses was analyzed [9].

Present work is a continuation of the our previous studies extended for the nonlinear optical properties of the $Ag_{0.05}Ga_{0.05}Ge_{0.95}S_2eEr_2S_3$ glasses and focused to determine the mechanism of photoluminescence emission in the red and infrared spectral range (660–1600 nm) under blue diode laser light excitation (457 nm).

The SHG and THG measurements have been performed in order to find a promising material for the multi-functional optoelectronic devices. The corresponding applications will include both coherent doubled and tripled laser beams as well as non-coherent fluorescence signal for light emission. As a consequence one can fabricate a devices which simultaneously has the optical triggering as well as laser diode features. At the same the fluorescence properties of the Er$^{3+}$ are not directly connected with the observed nonlinear optical response. The role of the Er ions was mainly caused by additional polarization of the media and changes of the effective hyperpolarizabilities defining the nonlinear optical constant due to interactions with the localized Er levels.

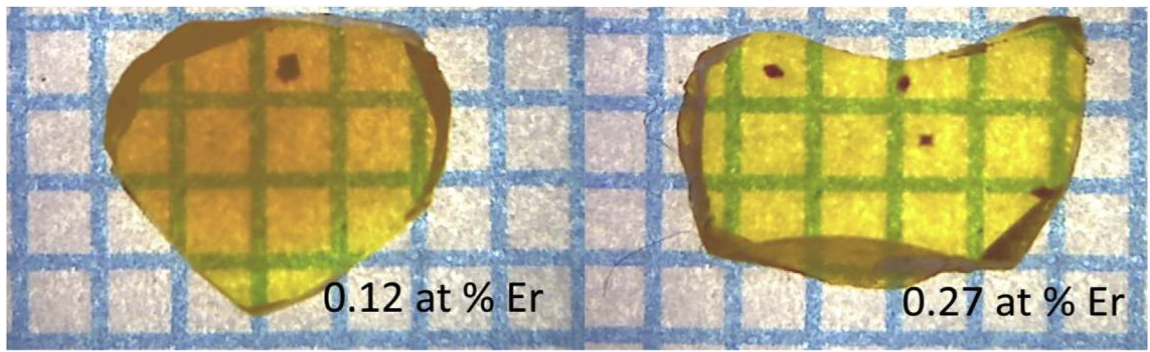

Fig. 1. Photographs of the glassy samples Ag$_{0.05}$Ga$_{0.05}$Ge$_{0.95}$S$_2$eEr$_2$S$_3$. Scale: 1 cell – 1 mm.

## 2. Experimental

The glasses were synthesized from the elementary substances (Ag, Ga, Ge – 99.997 wt%, S – 99.999% wt.%, Er – 99.9 % wt.%) in evac-uated quartz ampoules (residual pressure 0.1 Pa). To prevent con-densation losses of the vapor phase, the free volume of the container was wrapped by the asbestos cord. The batch was melted in the oxygen-gas burner flame to complete binding of the elementary sulfur; other-wise, its high vapor pressure could cause an explosion. Then, the am-poules were placed in a shaft-type furnace and heated at the rate of 20 K/h to 1273 K. The samples were kept at this temperature for 10 h, and then quenched by immersion into iced 25% aqueous NaCl solution. Photographs of the obtained samples are shown in Fig. 1. The glassy state of alloys was monitored by X-ray diffractometer with CuKα ra-diation (Fig. 2). The chemical composition of the prepared glasses was determined by energy-dispersive X-ray spectroscopy (EDS) analysis (Oxford X-Max 80 mm$^2$ module attached to a Tescan Mira 3 LMU SEM instrument).

The PL spectra were performed by using an MDR-24 mono-chromator as a spectrometer. The PL in the spectral range from 450 nm to 1100 nm was recorded using the CCD matrix (Andor), and in the region from 1100 to 1600 nm by Ge photodiode. The correction of the experimental spectra to the sensitivity of the instrument was applied in both spectral ranges, Laser with 457 nm and a power of ~10 mW was used for the excitation. The radius of the focused beam on the sample surface was 10 μm.

## 3. Results and discussion

### 3.1. Structure and composition

Commonly, the GeS$_2$ is used as a glass matrix for the incorporation of rare earth atoms. This is due to its optical characteristics, the wide transparency window in the visible and infrared spectral regions and high refractive index, in particular. Incorporation of Ga to GeS$_2$ extends the glass formation area, and also increases solubility of rare earth elements in the host matrix.

Due to mutual interactions between glass matrix and RE dopant, the energy levels in the 4f-shell of the erbium ions are split into a totality of sublevels (Stark splitting), thus affecting both the spectral behavior and the intensity of the photoluminescence. Adding a small percentage of Ag atoms to the compound forming glass matrix increases the efficiency of the photoluminescence of the incorporated Er$^{3+}$ ions. Consequently, the specimens, grown by the method described above, have yellow color (Fig. 1) and are transparent in a wide spectral range [5].

Amorphous nature of the glass matrixes for both samples was ver-ified using the X-ray diffraction analysis (Fig. 2). Using the EDS diag-nostic (Fig. 3), the composition of the synthesized glasses was de-termined, which well are agreed with the percentage content of the elementary components were used for synthesis.

### 3.2. Photoluminescence

The PL measurements of the Ag$_{0.05}$Ga$_{0.05}$Ge$_{0.95}$S$_2$eEr$_2$S$_3$ glasses with Erbium content 0.12 and 0.27 at.% were performed at room temperature and shown in Fig. 4. Four radiation bands in visible and near infrared spectral range with maxima at 660, 805, 980 and 1540 nm were recorded. The emission scheme is detailly presented in the diagram of energy levels of the Er$^{3+}$ ions (Fig. 5). Laser irradiation wavelength at 457 nm excites valence electrons of the erbium ions from the $^4I_{15/2}$ ground state to the state $^4F_{5/2,\ 3/2}$. The energy gap between the $^4F_{5/2,\ 3/2} - {}^4F_{7/2}$ and $^4F_{7/2} - {}^2H_{11/2}$ states is [18] 1700 and 1300 cm$^{-1}$, respectively. According to our previous studies [17], the energy phonons of glasses Ag$_{0,05}$Ga$_{0,05}$Ge$_{0,95}$S$_2$eEr$_2$S$_3$ is about 350 cm$^{-1}$. Thus, to transition Erbium ions from the state to $^4F_{5/2,\ 3/2}$ in the lower states ($^4F_{7/2}$ or $^2H_{11/2}$), it is necessary to radiate 4–5 phonons, which is unlikely. For comparison, such non-radiation transitions are possible in oxide glasses, in which the energy of phonons is about 700–800 cm$^{-1}$ [19]. This is the advantage of our glasses – small energy losses and a low probability of non-radiation relaxation. Therefore, the formation of excited states $^4F_{7/2}$, $^2H_{11/2}$, of which FL (980, 805 nm) occurs, due to non-radiation relaxation in glasses $Ag_{0.05}Ga_{0.05}Ge_{0.95}S_2eEr_2S_3$ is unlikely. The realization of these states in the $Er^{3+}$ ions is due to cross-relaxation. This process marked as CR1 leads to excited state $^4F_{7/2}$, which results in a powerful PL band with a maximum at 980 nm:

$$^4F_{5/2,\,3/2} + {}^2H_{11/2} \rightarrow 2\,{}^4F_{7/2} \tag{1}$$

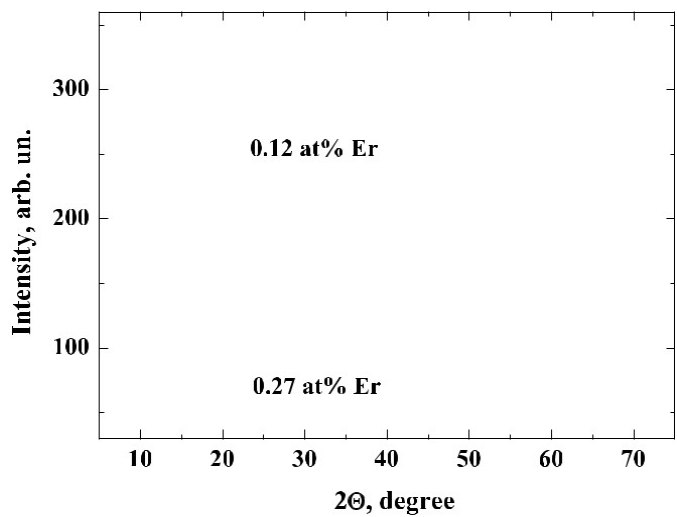

Fig. 2. X-ray diffraction patterns of the Ag$_{0.05}$Ga$_{0.05}$Ge$_{0.95}$S$_2$eEr$_2$S$_3$ glasses.

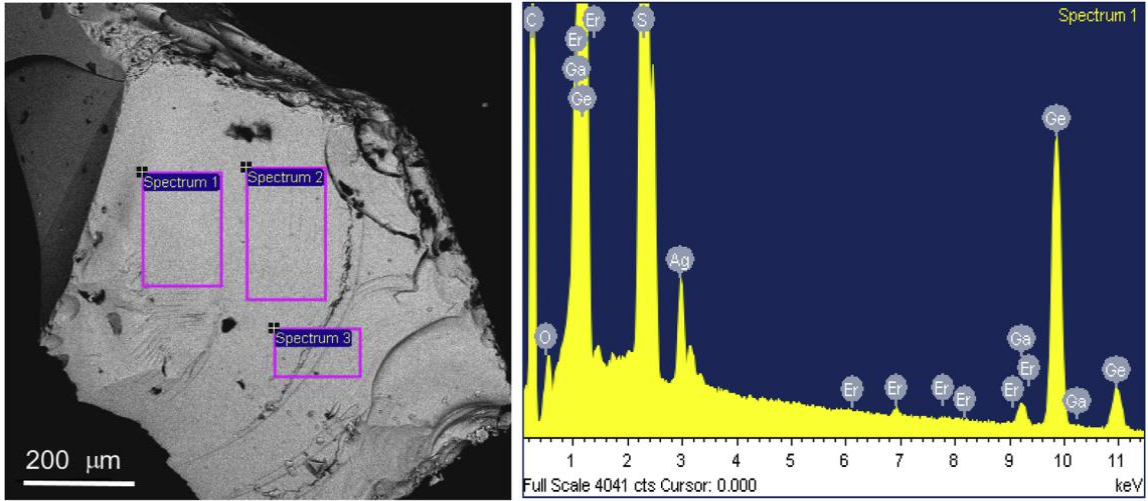

Fig. 3. SEM micrographs and experimental EDS analysis of the glasses $Ag_{0.05}Ga_{0.05}Ge_{0.95}S_2eEr_2S_3$ (0.27, ат.% Er).

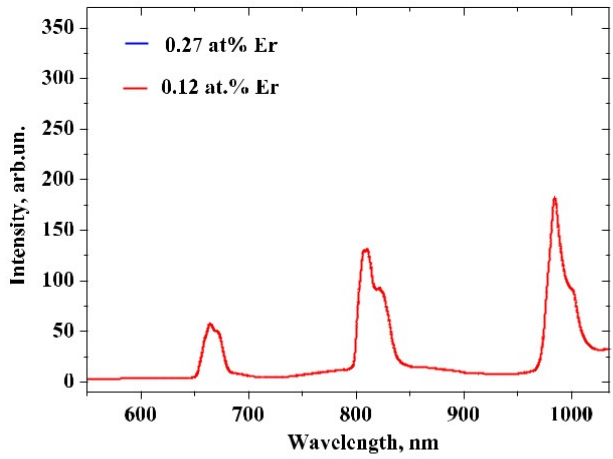

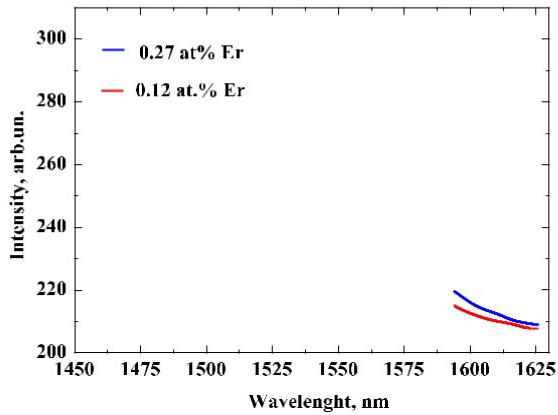

Fig. 4. Room-temperature PL of the $Ag_{0.05}Ga_{0.05}Ge_{0.95}S_2eEr_2S_3$ glasses in the spectral range 450–1100 nm (a) and 1450–1625 nm (b).

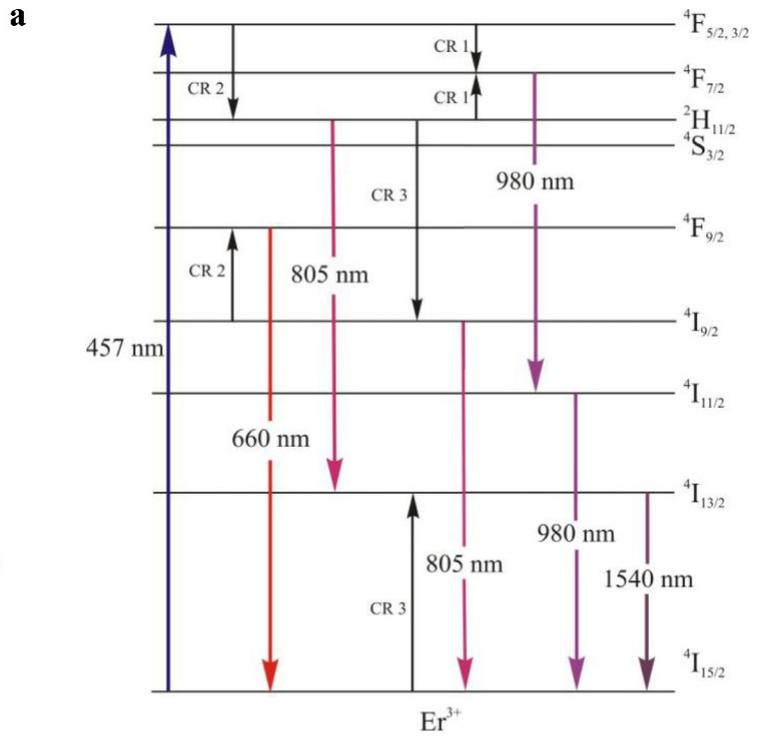

Fig. 5. Diagram of energy levels in $Er^{3+}$ ions.

Additionally, cross-relaxation CR2 yields excited states $^2H_{11/2}$ and $^4F_{9/2}$, resulting in emission with maxima at 805 and 660 nm:

$$^4F_{5/2,\,3/2} + {}^4I_{9/2} \rightarrow {}^2H_{11/2} + {}^4F_{9/2} \tag{2}$$

The high intensity of infrared PL with a maximum at 1540 nm is related to the fact that the state $^4I_{13/2}$ resulting the radiative transition $^2H_{11/2} \rightarrow {}^4I_{13/2}$ (maximum at 805 nm), as well as due to cross-relaxation CR3:

$$^2H_{11/2} + {}^4I_{15/2} \rightarrow {}^4I_{9/2} + {}^4I_{13/2} \tag{3}$$

It should be noted that for higher erbium content, the intensity of the band 1540 nm increases the least. The ratio of integral intensities for the band 1540 nm is $I_{0.27}/I_{0.12} = 1.03$, whereas for other bands the ratio is $I_{0.27/0.12} = 1.68$ (980 nm), $I_{0.27/0.12} = 2.15$ (805 nm), $I_{0.27/}$

$_{0.12}$ = 4.12 (660 nm). Calculations shown that for the higher Erbium concentration, the efficiency PL in the visible spectral range increases significantly and in contrast to this, the luminescence efficiency in the infrared spectral range only slightly increased. This interesting phenomenon can be explained as follows: the PL band with the maximum at 660 nm appears only due to CR2, while PL at 1540 nm results from the excited state $^4I_{13/2}$ which can be realized in two ways (Fig. 5): 1) involving cross-relaxation processes CR2 and radiative transition with the maximum at 805 nm; 2) cross-relaxation processes CR2 and CR3. The probability of a process that is due to the occurrence of two other processes will always be less than the one-act process. This causes a small gain in the intensity of the PL band with a maximum of 1540 nm compared with strips characterized by a shorter wavelength.

Therefore, an important role in the emission mechanism, manifested in the $Ag_{0.05}Ga_{0.05}Ge_{0.95}S_2eEr_2S_3$ glasses play the CR processes be-tween adjacent $Er^{3+}$ ions. Such process of energy transfer is char-acteristic for neighboring erbium ions which are grouped into clusters [20]. Through complementary EPR and static magnetization studies [9] has been explained, that some part of erbium ions in the $Ag_{0.05}Ga_{0.05}Ge_{0.95}S_2eEr_2S_3$ glasses is evenly distributed in the glass-forming matrix, while the other is involved in the formation of clusters that contain up to $10^3$ erbium ions. These glasses (in comparison with other Er-doped materials) are exceptional by the fact, that at 457 nm excitation, all of the PL bands are intense.

### 3.3. Non-linear optical properties

The photoinduced second (SHG) and third harmonic (THG) generation intensities have been measured in the reflected geometry. Measurements were repeated in different points of the studied samples to minimize impact of the possible local inhomogeneities, as well as the quality of their surface. As a fundamental laser, we have used the 1064 nm Nd:YAG 10 ns laser with frequency repletion of about 10 Hz. The beam diameter was equal to about 2 nm and the power density has been varied using the Glahn rotating polarizers. The samples have been investigated using successive measurements of the photoinduced THG like for the photoinduced SHG [21]. As the reference material, we have used well-studied oxide glasses containing lead [22,23]. The angle of incidence was changed continuously in the of 18–21° range to reach the highest SHG and THG intensity values. All samples along with reference sample (BBO:10%Nd crystal) were arranged between the shut-off in-terferometric filters (at 532 nm and 335 nm for second and third harmonic generation respectively). The photomultipliers have been connected with the PC computer across a Tektronix oscilloscope with 1 GHz sampling. Additionally, the scattered fundamental light was controlled. The typical THG dependences of such statistically averaged signals are presented at the Fig. 6. It is crucial, that with increasing Er content the output THG is also increased.

The dependence of the SHG and THG intensities versus the fundamental laser beams is presented in the Figs. 6 and 7. The efficiency of the SHG with respect to the extrapolated angle dependences was compared with the reference $BiB_3O_6$ Nd crystallites. It was appeared that the studied crystals have the efficiencies equal to about 16% with respect to the reference one. It is crucial that this dependence is almost opposite for the SHG and THG. This may be caused by different origin of the second and third order nonlinearities for the case of the Er doped samples. The second order hyperpolarizabilities are prevailingly de-termined by local charge density acentricity in the vicinity of the do-pants. And the higher order nonlinear optical effects mainly are de-tected by the excitation of the higher excited levels which differently interact with the ER localized state. Moreover, there is a significantly different dependence of the NLO constants versus the fundamental signal. The obtained signal for the THG is principally different due to realization of the different resonance conditions for the higher virtual states than for the SHG. It may confirm the mentioned different me-chanisms.

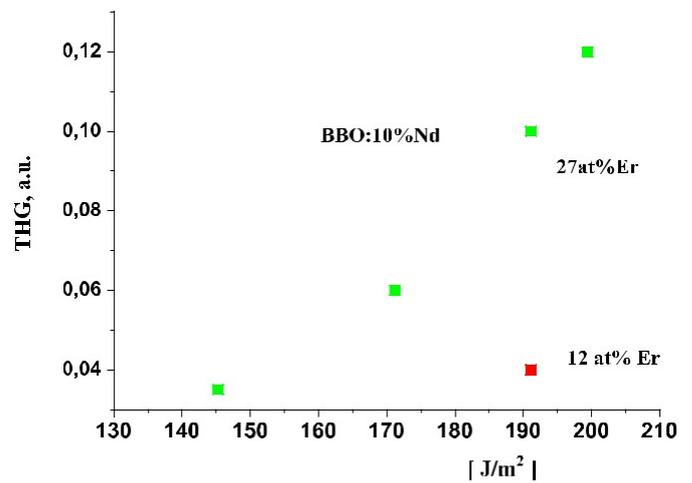

Fig. 6. The THG dependences for $Ag_{0.05}Ga_{0.05}Ge_{0.95}S_2eEr_2S_3$ glasses (red- with 0.12 and violet - 0.27 at.% Er, the green line reference THG for BBO:10%Nd crystal. (For interpretation of the references to color in this figure legend, the reader is referred to the Web version of this article.)

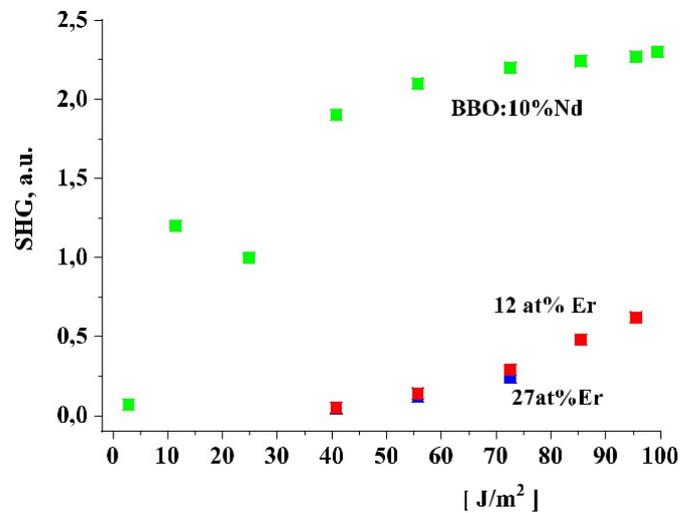

Fig. 7. The SHG dependences for $Ag_{0.05}Ga_{0.05}Ge_{0.95}S_2eEr_2S_3$ glasses (red- with 0.12 and violet - 0.27 at.% Er), the green line presents reference data of the THG intensity obtained for BBO:10%Nd crystal.

The increasing $Er^{3+}$ content from one side favors an enhanced local charge density a centricity which is responsible for the second order NLO susceptibilities like SHG. At the same time the THG is realized due to excitation of the higher excited states. So the space non-cen-trosymmetry does not play here a principal role and may be determined by the specific confutation of the $Er^{2+}$ ions.

Comparing the Figs. 6 and 7 defining the nonlinear optical properties and the linear optical features (Fig. 4) it is obvious that there exists some correlation between the second order NLO changes with the linear optical features like fluorescence. However, it is not observed for the third harmonic signal where the multi-photon excitation of the virtual states do not allow to find the direct correlation.

The principal difference with the previous articles [13] consists in a fact that we have simultaneously analyze both the SHG and THG and we demonstrate principally different behaviors on the Er dependent content.

### 4. Conclusions

The $Ag_{0.05}Ga_{0.05}Ge_{0.95}S_2eEr_2S_3$ glasses were synthesized by the two-stage method, and their glass condition and qualitative composition were checked by X-ray diffraction and EDS analysis, respectively. The conversion PL bands with the maxima at 660, 805, 980 and 1540 nm were recorded at room temperature under excitation by a 457 nm laser. With higher Erbium content, the intensity of the PL bands with maxima 660, 805, 980 nm greatly increases, but at 1540 nm we record a slight increase of intensity. This is due to the fact that the FL at 1540 is related to the existence of excited states $^4I_{13/2}$, which arise from two successive processes (cross-relaxation and PL radiation or two cross-relaxation processes). Performed investigations of the nonlinear optical properties indicate the inverse dependence of photoinduced SHG and THG intensity depending on erbium content. It is worth emphasizing the high efficiency of THG generation compared to the well-known BiBO:Nd nonlinear crystal.

Acknowledgements

Concerning K.O, and J.J. presented results are part of a project that has received funding from the European Union's Horizon 2020 research and innovation programme under the Marie Skłodowska-Curie grant agreement No 778156.